\documentclass[twocolumn,prd,groupedaddress,nofootinbib]{revtex4}
\usepackage{amssymb}
\usepackage{latexsym}

\begin{document}

\title{Corrections to the fine structure constant in higher dimensional
global monopole spacetime}
\author{Geusa de A. Marques$^{1\ast }$and V. B. Bezerra$^{2\ast \ast }$.}

\date{\today }

\begin{abstract}
In this paper we use the Generalized Uncertainty Principle(GUP) to
obtain the corrections to the fine structure constant in $(D+1)$-dimensional
global monopole spacetime. The result is particularized to $D$-dimensional
spacetime. We also discuss the particular case $D=3$ corresponding to the $%
(3+1)$-dimensional global monopole spacetime.

\smallskip
\end{abstract}
\affiliation{$^{1}$Departamento de F\'{\i}sica, Universidade Federal de Campina Grande\\
58109-790, Campina Grande, Pb, Brazil.\\
\smallskip \\
$^{2}$Departamento de F\'{\i}sica, Universidade Federal da Para\'{\i}ba\\
58051-970, Jo\~{a}o Pessoa, Pb, Brazil.}
\maketitle

\section{Introduction}

The Standard Cosmological Model predicts that the Universe has experienced a
chain of phase transitions with spontaneous symmetry breaking. During this
process the so called topological defects arised, as for example, monopoles,
strings and domain walls\cite{2}. In particular, global monopoles appear in
models with broken global $SO(3)$ symmetry\cite{BarVil89}. The model
consists of the Higgs triplet of scalar fields with Lagrangian 
\begin{equation}
\mathcal{L}=-\frac{1}{2} \partial _{\mu }\phi ^{a}\partial ^{\mu }\phi ^{a}-%
\frac{1}{4}\lambda \left( \phi ^{a}\phi ^{a}-\eta ^{2}\right) ^{2},
\end{equation}%
where $a=1,2,3.$

The gravitational field corresponding to a global monopole is represented by
the following spherically symmetric line element\cite{BarVil89}
\begin{equation}
ds^{2}=-B(r)dt^{2}+A(r)dr^{2}+r^{2}\left( d\theta ^{2}+\sin ^{2}\theta
d\varphi ^{2}\right) ,  \label{Mon}
\end{equation}%
with the functions $A$ and $B$ given by 
\begin{equation}
B=A^{-1}=1-8\pi \frac{G \eta ^{2}}{c^4}-\frac{2GM(r)}{rc^2},
\end{equation}
where the parameter $\eta ^{2}$ is related with the linear energy density.

The scale of the monopole's core is defined by the Compton wavelength of
Higgs bosons, $\varrho \sim \lambda ^{-1/2}\eta ^{-1}$. The parameter $%
M=\lim_{r\rightarrow \infty }M(r)$ which characterizes the monopole's mass,
assumes values in the interval $0\leq 8\pi \eta ^{2}<1 $ and is given,
approximately, by\cite{HarLou90} 
\begin{equation}
M\approx -\frac{6\pi \eta }{\sqrt{\lambda }}.  \label{Mass}
\end{equation}

Therefore, asymptotically the spacetime of a global monopole corresponds to
the Schwarzschild spacetime with negative mass and an additional solid angle
deficit $\frac{32\pi ^{2} G \eta ^{2}}{c^4}$. The spacetime of a point-like monopole 
is obtained from the above metric by disregarding the internal structure of the
global monopole but preserving the solid angle deficit. It is described by
the following line element 
\begin{equation}
ds^{2}=-c^{2}b_{3}^{2}dt^{2}+ \frac{dr^{2}}{b_{3}^{2}}+r^{2}\left( d\theta
^{2}+\sin ^{2}\theta d\varphi ^{2}\right) ,  \label{GlobMon}
\end{equation}%
where $b_{3}^{2}=1-\frac{8\pi G \eta ^{2}}{c^4}$, with the
subindex indicating the dimension of space. The scalar curvature is
given by $R=\frac{1-b_{3}^{2}}{r^{2}b_{3}^{2}}$. Therefore, this spacetime
is not flat, but surprisingly there is no Newtonian potential associated to
it.

This metric can be written in a different form if we make the following
change in time coordinate $t\rightarrow t/b_{3}$. As a result, it turns into

\begin{equation}
ds^{2}=-c^{2}dt^{2}+\frac{dr^{2}}{b_{3}^{2}}+ r^{2}\left( d\theta ^{2}+\sin
^{2}\theta d\varphi^{2}\right).  \label{GlobMonMetric}
\end{equation}

The main characteristic of this spacetime is the existence of a solid angle
deficit $4\pi (1-b_{3}^{2})$ whose typical value in the framework of Grand
Unified Theory(GUT)($\eta ^{2}c^{2}L_{pl}\sim 10^{16}GeV$) is
proportional $\frac{8\pi G\eta ^{2}}{c^4}
\sim 10^{-5}$.

Taking into account that our proposal is to consider the higher dimensional
global monopole spacetime, let us write the generalization of the metric
given by eq.(\ref{GlobMonMetric}), for a $(D+1)$-dimensional spacetime. In
this case, the Euclidean version of the global monopole line element is
given by\cite{Eug} 

\begin{equation}
ds^{2} = d\tau^{2}+ \frac{dr^{2}}{b_{D}^{2}} + r^{2} d\Omega _{D-1}^{2},
\label{1}
\end{equation}
where $D\geq 3$ and $b_D$ is a parameter which codifies the presence of a $%
(D+1)$-dimensional global monopole. The coordinates are $(\tau, r, \theta_1,
\theta_2,..., \theta_{D-2}, \phi).$ In this case, the solid angle associated
to a hypersphere with unity radius is $\Omega_{b_D}=b_{D}^{2}\frac{2\pi
^{D/2}}{\Gamma \left( \frac{D}{2}\right)}$ which is smaller than the solid
angle associated to a hypersphere embbeded in flat spacetime.

\section{Corrections to the fine structure constant in higher dimensions}

In order to obtain the fine structure constant in the $(D+1)$-dimensional
global monopole spacetime, let us consider a Hydrogen atom in this
background. Thus, the electrostatic force on the electron is given by 

\begin{equation}
\vec{F}_{totD}=\left( -\frac{1}{\Omega _{D-1}\epsilon _{D,0}}\frac{e^{2}}{%
r^{D-1}}+\frac{S(b_{D}}{\Omega _{D-1}\epsilon _{D,0}}\frac{e^{2}}{ r^{D-1}}%
\right) \hat{r}_{D}, \label{autoforca} 
\end{equation}
where the first term corresponds to the generalization of the usual Coulomb
force to a $D$-dimensional space. The second term arises from the geometrical
and topological features of the spacetime and corresponds to a
generalization of a result obtained in the framework of a $(3+1)$-dimensional 
global monopole\cite{eugenio}. The factor $S(b_{D})$ is also a
generalization of a previous result\cite{eugenio} and is given by

\begin{equation}
S(b_{D})=\frac{1}{2}\sum_{n=1}^{\infty }\frac{\left( \pi ^{2}\Delta
_{D}\right)^n } {\left( n!\right) ^{2}}\left\vert B_{2n}\right\vert \left(
1-2^{-2n}\right) \label{autoforca2}
\end{equation}
with $\Delta _{D}=1-b_{D}^{2}$ and $B_{n}$ being the Bernoulli numbers.
Thus, we can obtain the Bohr radius in this background which can be written
as

\begin{eqnarray}
{r}_{B,D}&=&\left( \frac{b_{D}^{2}\Omega _{D-1}\epsilon _{D,0}\hbar ^{2}}{%
me^{2}}\right) ^{\frac{1}{4-D}} \nonumber  \\
&&\times \left( 1-S(b_{D})\right)^{4-D},  
\label{Bohrradius}
\end{eqnarray}
where $m$ and $e$ are the mass and charge of the electron, respectively.
Note that when $D=3$ and $b_{3}=1$, which means that the global monopole is
absent,we obtain the so-called Bohr radius $r_{B,3}=\frac{4\pi \epsilon _{0}\hbar
^{2}}{me^{2}}=0.0529 nm$ in three dimensional space. This expression for the
Bohr radius takes into account that the usual quantization of the angular
momentum is now given by $L_{n,D}= \frac{n \hbar}{b_D}$.

The fine structure constant in $D$-dimensional space was obtained recently\cite{NGUP} and is given by the following expression 
\begin{equation}
\alpha _{D}=\hbar ^{2-D}e^{D-1}[\Omega _{D-1}\epsilon _{D,0}]^{\left(
1-D\right) /2}c^{D-4}G_{D}^{(3-D)/2}  \label{4}
\end{equation}%
where $\epsilon _{D,0}$ is the permittivity constant and $\Omega _{D-1}=%
\frac{2\pi ^{D/2}}{\Gamma \left( \frac{D}{2}\right) }$ is the solid angle
associated with a hypersphere in $D$ dimensions. From this result, we can
obtain a corresponding generalized one in the sense that the global monopole
is present. This can be written as

\begin{equation}
\hat{\alpha}_{D}=\hbar ^{2-D}e^{D-1}[b_{D}^{2}\Omega _{D-1}\epsilon
_{D,0}]^{\left( 1-D\right) /2}c^{D-4}G_{D}^{(3-D)/2},  \label{5}
\end{equation}
where the parameter $b_D^{2}$ was introduced to take into account the
presence of the $(D+1)$-dimensional global monopole. For $D=3$ and $b_{3}=1$, eq.(\ref{5}) turns into $\alpha=e^{2}/4\pi \epsilon \hbar c \approx 1/137$

The so-called Generalized Uncertainty Principle\cite{maggiore} is
expected to be satisfied when quantum gravitational effects are important.
The interest in this Principle has been motivated by studies in string theory\cite{gross} and gravity\cite{maggiore}. In this context, a minimal length
scale is expected to be of the order of Planck length $L_p$. This fact leads
to corrections to the usual Heisenberg Uncertainty Principle, resulting in
the Generalized Uncertainty Principle, which can be expressed as

\begin{equation}
\Delta x_{i}\Delta p_{i}\geq \hbar \left[ 1+\beta ^{2}L_{p,D}^{2}\left( 
\frac{\Delta p_{i}}{\hbar }\right) ^{2}\right] .  \label{10}
\end{equation}
where $\beta $ is a numerical factor of order unity that depends on the
specific model and $L_{p,D}$ is the Planck length in $D$ dimensions, which
is given by $L_{p,D}=(\hbar G_{D}/c^{3})^{1/D-1}$. Thus, in a three
dimensional space eq.(\ref{10}) implies that the lower bound on the length
scale is of the order of Planck's length, $L_p = \sqrt(G\hbar/c^3)\approx 10^{-33}cm$, 
and this plays an important role as a fundamental scale.

Thus, from eq.(\ref{10}) we can obtain the uncertainty for the momentum
which is given by the following relation 
\begin{equation}
\frac{\Delta p_{i}}{\hbar }=\frac{\Delta x_{i}}{2\beta L_{p,D}^{2}}\left( 1-%
\sqrt{1-\frac{4\beta ^{2}L_{p,D}^{2}}{\Delta x_{i}^{2}}}\right) .  \label{9}
\end{equation}

Now, let us consider that the uncertainty in position is equal to the Bohr
radius given by eq.(\ref{Bohrradius}). Thus, we can write the following relation

\begin{eqnarray}
\frac{\Delta x_{i}}{L_{p,D}} &=&\left( \frac{b_{D}^{2}\Omega _{D-1}\epsilon
_{D,0}M_{p,D}^{3}G_{D}}{me^{2}}\right) ^{\frac{1}{4-D}} \nonumber \\ 
&&\times \left( 1-S(b_{D})\right) ^{4-D},  \label{9b}
\end{eqnarray}
where $M_{p,D}=(\hbar^{D-2}/c^{D-4}G_D)^{1/D-1}$is the Planck's mass in $D$ 
dimensional space.

Comparing the Generalized Uncertainty Principle and the Heisenberg
Uncertainty Principle, we conclude that the first one can be written, formally, as

\begin{equation}
\Delta x_{i} \Delta p_{i} \geq \hbar_{eff},  \label{VBB1}
\end{equation}
where 
\begin{equation}
\hbar _{eff}=\hbar \left[ 1+\beta ^{2}L_{p,D}^{2} \left( \frac{\Delta p_{i}}{%
\hbar }\right) ^{2} \right] .  \label{11}
\end{equation}

Substituting eq.(\ref{9b}) into (\ref{9}) and the result into (\ref{11}),
we get 

\begin{widetext}

\begin{equation}
\hbar _{eff}=
\begin{array}{c}
\hbar \left[ 1+\frac{1}{4\beta ^{2}}\left( \frac{b_{D}^{2}\Omega
_{D-1}\epsilon _{D,0}M_{p,D}^{3}G_{D}}{me^{2}}\right) ^{\frac{2}{4-D}%
}(1-S(b_{D}))^{2(4-D)}\right.  \\ 
\times \left. \left( 1-\sqrt{1-4\beta ^{2}\left( \frac{me^{2}}{%
b_{D}^{2}\Omega _{D-1}(1-S(b_{D}))^{2(4-D)}}\epsilon
_{D,0}M_{pD}^{3}G_{D}\right) ^{\frac{2}{4-D}}}\right) ^{2}\right] 
\end{array}.  \label{hefetivo}
\end{equation}

\end{widetext}

If we consider a global monopole in the Grand Unified Theory framework,
the factor $\frac{b_D}{(1-S_D)}$ should be approximately equal to unity, as
we can see for the particular case $D=3$, in which situation we have $%
b_3=0.9999$, $S(b_3)=0.0024$ and $\frac{1}{(1-S_3)^{2}b_{3}^{4}} \approx 1$.
Therefore, we can use the term $\frac{me^{2}}\Omega_{D-1}\epsilon
_{D,0}M_{p,D}^{3}G_{D}$,which is much less than one, as the expansion parameter.
Thus, we can expand eq.(\ref{hefetivo}) in terms of this quantity, which gives us
the following result 
\begin{eqnarray}
\hbar _{eff} &\simeq &\hbar \left[ 1+\beta ^{2}\left( \frac{me^{2}}{
b_{D}^{2}\Omega _{D-1}\epsilon _{D,0}M_{p,D}^{3}G_{D}}\right) ^{\frac{2}{4-D}
}\right.  \label{12} \\
&&\left. \times \left( 1-S(b_{D})\right) ^{-2(4-D)} \right] .  \nonumber
\end{eqnarray}

Note that in the absence of a global monopole, the result given by eq.(\ref%
{12}) corresponds to the result obtained by Nasseri\cite{NGUP} in a $D$-dimensional space. Applying this result given by eq.(\ref{12}) to a $(3+1)$-dimensional global monopole spacetime, we get 
\begin{equation}
\hbar _{eff} \simeq \hbar ( 1+ 9.53 \times 10^{-50} \beta ^{2} ).
\label{VBB2}
\end{equation}

Combining eqs.(\ref{5}) and (\ref{12}), we find that the structure constant
in a $(D+1)$-dimensional global monopole spacetime is given by the expression 
\begin{equation}
\hat{\alpha}_{D,eff}=\hbar _{eff}^{2-D}e^{D-1}[b_{D}^{2}\Omega
_{D-1}\epsilon _{D,0}]^{\left( 1-D\right) /2}c^{D-4}G_{D}^{(3-D)/2},
\end{equation}
which reduces to
\begin{equation}
\hat{\alpha}_{3,eff}=\frac{e^{2}}{4\pi \epsilon _{0} \hbar c}(1 - 9.53 \times
10^{-50} \beta^2),
\end{equation}
in a $(3+1)$-dimensional global monopole spacetime.
This result shows that the value of the fine structure constant is very close 
to the one obtained in three-dimensional Minkowski flat space. 

\section{Concluding remarks}

The corrections to the Planck constant and to the fine structure constant
are very close to the ones obtained in a $D$-dimensional flat spacetime. The
difference between $\hbar_{eff}$ and $\hbar$ in relation to $\hbar$ is of the
order of $10^{-49}$, when we consider the presence of a global monopole as
well as in the absence of this topological defect.
Therefore, the values of the fine structure constant
evaluated in the framework of the Generalized Uncertainty Principle and in
the presence of a $(D+1)$-dimensional global monopole or in the absence of
this topological defect, are very close to the value of the fine structure
constant obtained in the framework of the Heisenberg Uncertainty Principle.

{\large \textbf{Acknowledgments}} We would like to thank Conselho Nacional
de Desenvolvimento Cient\'{\i}fico e Tecnol\'ogico(CNPq),
FAPESQ-PB/CNPq-PRONEX and FAPES-ES/CNPq-PRONEX for the partial financial
support.

$^{\ast }$Electronic address: gmarques@df.ufcg.edu.br

$^{\ast \ast }$Electronic address: valdir@fisica.ufpb.br

\begin{thebibliography}{9}
\bibitem{2} A. Vilenkin and E. P. S. Shellard, {\it Cosmic string and
Other Topological Defects}, Cambridge Univ. Press, Cambridge, 1994.

\bibitem{BarVil89} Manuel Barriola and A. Vilenkin, Phys. Rev. Lett. {\bf
63}, 341 (1989).

\bibitem{HarLou90} D. Harari and C. Lousto Phys. Rev. D {\bf 42}, 2626
(1990).

\bibitem{Eug} E. R. Bezerra de Mello, J. Math. Phys. {\bf 43}, 1018 (2002).

\bibitem{eugenio} E. R. Bezerra de Mello and C. Furtado, Phys. Rev. D 
{\bf 56}, 1345 (1997).

\bibitem{NGUP} Forough Nasseri, Phys. Lett. B {\bf 618}, 229 (2005).

\bibitem{maggiore} M. Maggiore, Phys. Lett. B {\bf 304}, 65 (1993).

\bibitem{gross} D. J. Gross and P. F. Mendle, Nucl. Phys. B {\bf 303},
407 (1988); D. J. Gross, Phys. Rev. Lett. {\bf 60}, 1229 (1988); G. Amati,
M. Ciafaloni and G. Veneziano, Phys. Lett. B {\bf 216}, 41 (1989).
\end{thebibliography}
\end{document}